\documentclass[a4paper,3p]{elsarticle}
\biboptions{sort,compress}
\usepackage[T1]{fontenc}
\usepackage[utf8]{inputenc}
\usepackage{color}
\usepackage[english]{babel}
\usepackage{amsmath}
\usepackage{amssymb}
\usepackage{graphicx}
\usepackage[hidelinks]{hyperref}

\journal{Annals of Physics}

\makeatletter

\newcommand{\be}{\begin{equation}}
\newcommand{\ee}{\end{equation}}
\newcommand{\bea}{\begin{eqnarray}}
\newcommand{\eea}{\end{eqnarray}}
\newcommand{\bse}{\begin{subequations}}
\newcommand{\ese}{\end{subequations}}

\def\bmx{\begin{pmatrix}}
\def\emx{\end{pmatrix}}

\makeatother

\begin{document}

\title{Eliashberg equations for an electron-phonon version of the Sachdev-Ye-Kitaev
model: Pair Breaking in non-Fermi liquid superconductors}

\author[1]{Daniel Hauck}
\author[1]{Markus J. Klug}
\author[2]{Ilya Esterlis}
\author[1,3]{Jörg Schmalian}

\address[1]{Institute for Theory of Condensed Matter, Karlsruhe Institute of
Technology, Karlsruhe 76131, Germany}
\address[2]{Department of Physics, Harvard University, Cambridge, Massachusetts
02138, USA}
\address[3]{Institute for Quantum Materials and Technologies, Karlsruhe Institute of Technology,
Karlsruhe 76021, Germany}

\begin{abstract}
We present a theory that is a non-Fermi-liquid counterpart of the
Abrikosov-Gor\textquoteright kov pair-breaking theory due to paramagnetic
impurities in superconductors. To this end we analyze a model of interacting
electrons and phonons that is a natural generalization of the Sachdev-Ye-Kitaev-model.
In the limit of large numbers of degrees of freedom, the Eliashberg
equations of superconductivity become exact and emerge as saddle-point
equations of a field theory with fluctuating pairing fields. In its
normal state the model is governed by two non-Fermi liquid fixed points,
characterized by distinct universal exponents. At low temperatures
a superconducting state emerges from the critical normal state. We
study the role of pair-breaking on $T_{c}$, where we allow for
disorder that breaks time-reversal symmetry. For small Bogoliubov
quasi-particle weight, relevant for systems with strongly incoherent
normal state, $T_{c}$ drops rapidly as function of the pair breaking
strength and reaches a small but finite value before it vanishes at
a critical pair-breaking strength via an essential singularity. The
latter signals a breakdown of the emergent conformal symmetry of
the non-Fermi liquid normal state.
\end{abstract}

\begin{keyword}
    Eliashberg Theory\sep Superconductivity\sep Non-Fermi Liquid\sep
    Sachdev-Ye-Kitaev Model\sep Pair Breaking
\end{keyword}

\maketitle

\section{Introduction}

The dynamical theory of phonon-mediated superconductivity was formulated by Gerasim Matveevich Eliashberg in a pioneering tour de force of quantum many-body theory\cite{Eliashberg1960,Eliashberg1961}.
Considering the regime where phonon frequencies are much smaller than
the Fermi energy of the electrons, electrons follow the lattice motion
almost instantly. In this limit, Migdal had shown that electron-phonon
vertex corrections become small\cite{Migdal1958}. Then a complicated
intermediate-coupling problem suddenly becomes tractable. A closed,
self-consistent dynamical theory emerges that is not limited to the
regime of weak electron-phonon interactions. The Eliashberg formalism
follows the Gor\textquoteright kov-Nambu description of superconductivity\cite{Gorkov1958,Nambu1960},
reflecting the broken global $U\left(1\right)$ symmetry, associated
with charge conservation. The propagation of particles and the conversion
of particles into holes are described by two self energies $\Sigma\left(\omega\right)$
and $\Phi\left(\omega\right)$, respectively.
Using the Eliashberg
theory, important advances were made in understanding the physical properties
of superconductors with a dimensionless electron-phonon coupling of
order unity\cite{Schrieffer1963,Scalapino1966,McMillan1968,Scalapino1969,McMillan1969,Allen1975,Carbotte1990}.

\begin{figure}[t!]
    \centering
\includegraphics[width=\textwidth]{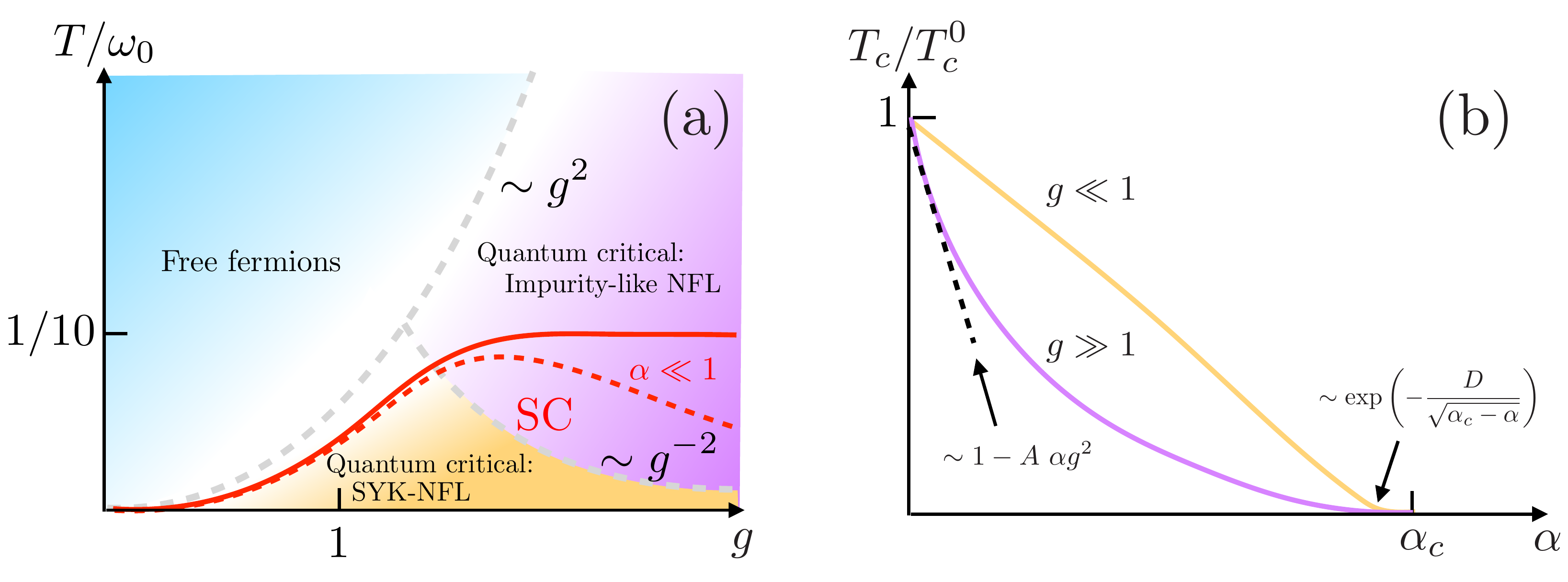}

\caption{(a) Schematic phase diagram obtained from solving the Eliashberg equations, Eq.\eqref{eq:Eliash_eqs}. Normal state properties are independent of pair breaking parameter $\alpha$. The solid red line is superconducting $T_c$ for $\alpha = 0$; dashed red line is $T_c$ for a small non-zero value of $\alpha \ll 1$. For $\alpha$ small, $T_c$ is weakly affected at weak coupling $g \ll 1$. However, for strong coupling $g \gg 1$ even small $\alpha$ dramatically depresses $T_c$. (b) Schematic plot of superconducting $T_c$ as a function of pair breaking parameter $\alpha$, normalized by $T_c^0 = T_c(\alpha = 0)$. In the weak coupling regime $T_c(\alpha)/T_c^0$ follows a $g$-independent universal curve. In the strong coupling regime $T_c$ drops linearly with $\alpha g^2$ for small $\alpha$, so that $T_c$ is already significantly reduced for $\alpha \sim 1/g^2$.  $T_c(\alpha)$ extends to a $g$-independent critical value $\alpha_c$. Near $\alpha_c$ the transition temperature obeys a BKT-scaling $T_c \sim \exp(-D/\sqrt{\alpha_c -\alpha})$.}

\label{fig:phase_diag}
\end{figure}

The Eliashberg formalism has been applied to study superconductivity
in problems that go significantly beyond the original electron-phonon
problem\cite{Bonesteel1996,Son1999,Abanov2001,Abanov2001b,Roussev2001,Chubukov2005,Metlitski2015,Raghu2015,Wang2016,Abanov2019,Wu2019}.
When an electronic system becomes quantum critical, soft degrees of
freedom emerge. The retarded nature of the coupling to such soft excitations makes
an analysis in the spirit Eliashberg's approach, with
a dynamical pairing field $\Phi\left(\omega\right)$, natural. Since
realistic models of quantum critical pairing usually possess no natural
small parameter, a controlled approach that leads to an Eliashberg-like
formalism is highly desirable.

Recently, two of us introduced and solved a model for the electron-phonon
interaction in non-Fermi liquids\cite{Esterlis2019}. It is a natural
generalization of the Sachdev-Ye-Kitaev (SYK) model\cite{Sachdev1993,Georges2000,Sachdev2010,Kitaev2015,Kitaev2015b}
and yields superconductivity due to electron-phonon interactions with
quantum critical behavior in the normal state. The model becomes solvable
in the limit of infinite number of degrees of freedom. The Eliashberg
equations of superconductivity, with self-consistently determined
electron and phonon propagators, become exact. The formalism
yields rich non-Fermi liquid behavior in the normal state
and gives rise to superconductivity at low temperatures. Related interesting
descriptions of superconductivity in SYK-like models have also been discussed
in Refs.\cite{Patel2018,Wang2019,Chowdhury2019}. While SYK models
are dominated by random interactions, the belief is that the non-Fermi
liquid behavior that occurs is in fact more general and may also offer
insights into non-random systems.

In its normal state the model of Ref.\cite{Esterlis2019} is governed
by two non-Fermi liquid fixed points, characterized by distinct universal
exponents. The weakened ability of such non-Fermi liquid electronic
states to form Cooper pairs is offset by an increasingly singular
pairing interaction, leading to coherent superconductivity in such
incoherent systems. This result is closely related to the generalized
Cooper theorem of quantum-critical pairing put forward by Abanov {\em et al.}  in Ref.\cite{Abanov2001}.
In Ref.\cite{Esterlis2019} the ground state was shown to be characterized
by sharp Bogoliubov quasiparticles. However, the incoherent nature of the normal-state leads to a much reduced spectral weight $Z_{B}$ of the Bogoliubov
quasiparticles. For small values of $Z_{B}$ a reduction in the condensation
energy occurs. At the same time, the transition temperature remains
unchanged. This behavior is reminiscent of superconductivity in systems
with non-pair-breaking impurities, where Anderson's theorem guarantees
an unchanged transition temperature, while the superconducting state
becomes more fragile the larger the disorder strength, with e.g. a
strongly reduced superfluid stiffness\cite{Anderson1959,Abrikosov1958,Abrikosov1959}.

An important issue in the investigation of superconducting states
is their robustness with respect to pair-breaking disorder. The topic
was pioneered by Abrikosov and Gor'kov, who analyzed the role of paramagnetic
impurities in conventional superconductors\cite{Abrikosov1961} and
found a suppression of $T_{c}$ determined by $\log\left(T_{c}^{0}/T_{c}\right)=\psi\left(\frac{1}{2}-\frac{1}{2\pi\tau T_{c}}\right)-\psi\left(\frac{1}{2}\right)$
with scattering rate due to paramagnetic impurities $\tau^{-1}$ and digamma function $\psi$.
$T_{c}^{0}$ is the transition temperature without pair breaking.
The inclusion of quantum dynamics of the impurities, unconventional
pairing states, critical normal states, and strong impurity scattering
are topics of ongoing theoretical investigations, see e.g.
Refs.\cite{ErwinMH1971,Yoksan1984,Monthoux1994,Preosti1996,Golubov1997,Franz1997,Haran1998,Kulic1999,Dzero2005,Balatsky2006,Graser2007,Alloul2009,Kemper2009,Kogan2009,Vorontsov2009,Vorontsov2010,Hoyer2015,Michaeli2012,Scheurer2015,Kang2016,Trevisan2018}
for an incomplete list of publications. In this context an interesting
question is the nature by which superconductivity vanishes
due to pair breaking if the normal state is quantum critical.

In this paper we generalize the electron-phonon SYK model to analyze
the robustness of pairing in quantum-critical systems against pair-breaking effects
due to time-reversal symmetry violation. We solve the modified Eliashberg
equations and find a suppression of the transition temperature as a function of
a pair-breaking parameter $\alpha$, with $T_{c}$ vanishing at a
critical pair-breaking strength $\alpha_{c}$. While the qualitative
trends are similar to the Abrikosov-Gor\textquoteright kov theory\cite{Abrikosov1961}, there are key distinctions
in the overall dependence of $T_{c}$ on $\alpha$. Near $\alpha_{c}\approx0.62$, we find a behavior
\begin{equation}
T_{c}\left(\alpha\approx\alpha_{c}\right) =T^* \exp\left(-\frac{D}{\sqrt{\alpha_{c}-\alpha}}\right),\label{eq:BKTscaling}
\end{equation}
where $D$ is a non-universal constant and $T^*$ an energy scale that we discuss below. This behavior is similar to
the scaling near a Berezinskii-Kosterlitz-Thouless (BKT) transition\cite{Berezinskii1972,Kosterlitz1973}.
Such BKT-scaling was argued to be generic for systems with a transition
from a conformal to a non-conformal phase\cite{Kaplan2009}. Given
the conformal symmetry of the SYK model\cite{Georges2000}, which
is relevant to our normal state, the result Eq.\eqref{eq:BKTscaling}
for the superconducting transition temperature is further confirmation
of the expectation put forward in Ref.\cite{Kaplan2009}. Thus, the
change of the superconducting transition temperature as function of
a pair-breaking impurity concentration may serve as a tool to identify
whether a normal state can be effectively thought of as a critical
state with an underlying conformal symmetry. A behavior like that of
Eq.\eqref{eq:BKTscaling} occurs in the coupling constant dependence of the mass
scale near the chiral symmetry breaking point of  $2+1$-dimensional quantum
electrodynamics\cite{Appelquist1988}. In the context of coherent versus
incoherent pairing, such behavior was first seen in an Eliashberg theory  near
a magnetic instability in Ref.\cite{Abanov2001}. An interesting renormalization
group perspective of Eq.\eqref{eq:BKTscaling} in the context of superconductivity was recently given in Ref.\cite{Raghu2015}.
An appeal of our approach is that the critical coupling $\alpha_c$ in Eq.\eqref{eq:BKTscaling} acquires a clear physical and potentially tunable interpretation as a pair-breaking parameter due to time-reversal symmetry breaking disorder.

In addition to the behavior near the critical pair-breaking strength
$\alpha_{c}$ we also analyze the interplay between normal-state incoherency
and the robustness of superconductivity with respect to pair breaking.
In the incoherent, strong coupling regime of the system, we find that
$T_{c}$ is already substantially suppressed for $\alpha\approx\alpha^{*}\ll\alpha_{c}$,
where the crossover scale $\alpha^{*}\sim Z_{B}$ is proportional
to the small Bogoliubov quasiparticle weight for $\alpha=0$, i.e.
\begin{equation}
T_{c}\left(\alpha\sim Z_{B}\right)\ll T_{c}^{0},\label{eq:strong coupling Tc'}
\end{equation}
with $T_{c}^{0}=T_{c}\left(\alpha=0\right)$. Thus, a pairing state that emerges from an incoherent normal state
with small $Z_{B}$ is particularly fragile against pair breaking. These findings are summarized in Fig.\ref{fig:phase_diag}b.

In what follows we introduce our model, show that the solution is
given by a set of coupled Eliashberg equations, and present the solution
of this set of equations.

\section{Eliashberg equations}

We start from the Hamiltonian
\begin{eqnarray}
H & = & -\mu \sum_{i=1}^{N}\sum_{\sigma=\pm} c_{i\sigma}^{\dagger}c_{i\sigma}+\frac{1}{2}\sum_{k=1}^{M}\left(\pi_{k}^{2}+\omega_{0}^{2}\phi_{k}^{2}\right)+\frac{1}{N}\sum_{ij,\sigma}^{N}\sum_{k}^{M}g_{ij,k}c_{i\sigma}^{\dagger}c_{j\sigma}\phi_{k},\label{eq:Hamiltonian}
\end{eqnarray}
with electron annihilation and creation operators $c_{i\sigma}$ and
$c_{i\sigma}^{\dagger}$, respectively, that obey $\left[c_{i\sigma},c_{j\sigma'}^{\dagger}\right]_{+}=\delta_{ij}$$\delta_{\sigma\sigma'}$
and $\left[c_{i\sigma},c_{j\sigma}\right]_{+}=0$ with spin $\sigma=\pm1$.
In addition we have phonons $\phi_{k}$ with canonical momentum $\pi_{k}$,
such that $\left[\phi_{k},\pi_{k'}\right]_{-}=i\delta_{kk'}$. Here
$i,j=1\cdots N$ refer to electrons and $k=1\cdots M$ to the
phonons. In what follows we mostly consider the limit $N=M$.

The problem becomes solvable because of the fully-connected nature
of the electron-phonon interaction. The electron-phonon coupling constants
$g_{ij,k}$ are Gaussian-distributed random variables that obey $g_{ij,k}=g_{ji,k}^{*}.$
The coupling constants are in general complex valued $g_{ij,k}=g_{ij,k}^{'}+ig_{ij,k}^{''}$
with real part $g_{ij,k}^{'}$ and imaginary part $g_{ij,k}^{''}$. In what follows we will use a distribution function
with zero mean and second moment given by
\begin{eqnarray}
\overline{g_{ij,k}^{'}g_{i'j',k'}^{'}} & = & \left(1-\frac{\alpha}{2}\right)\bar{g}^{2}\delta_{k,k'}\left(\delta_{ii'}\delta_{jj'}+\delta_{ij'}\delta_{ji'}\right),\nonumber \\
\overline{g_{ij,k}^{''}g_{i'j',k'}^{''}} & = & \frac{\alpha}{2}\bar{g}^{2}\delta_{k,k'}\left(\delta_{ii'}\delta_{jj'}-\delta_{ij'}\delta_{ji'}\right),\nonumber \\
\overline{g_{ij,k}^{'}g_{i'j',k'}^{''}} & = & 0.\label{eq:distribution}
\end{eqnarray}
The over-bar denotes disorder averages. The two limits $\alpha=0$
and $\alpha=1$ were discussed previously in Ref.\cite{Esterlis2019}.
In the former case, the coupling constants are all real valued with
$g_{ij,k}=g_{ji,k}$. For given $k$, the $g_{ij,k}$ are then chosen
from the Gaussian orthogonal ensemble of random matrices\cite{Mehta2004}.
Time-reversal symmetry of the Hamiltonian is not only preserved on
average, but also for each individual realization of the $g_{ij,k}$.
As a result, the electron-phonon interaction of Eq.\eqref{eq:Hamiltonian}
induces superconductivity. On the other hand, for $\alpha=1$, each
configuration of the coupling constants, distributed according to the Gaussian unitary ensemble,  strongly breaks time-reversal
symmetry and no superconductivity occurs. Instead, a non-Fermi liquid
normal state emerges where the spectral functions of electrons and
phonons are governed by universal power laws, a behavior that is qualitatively
similar to the usual formulation of the SYK model with random four-fermion
interactions\cite{Sachdev1993,Georges2000,Sachdev2010,Kitaev2015,Kitaev2015b}.
Below we will summarize the main finding of these limits in greater
detail.

For $0<\alpha<1$ the breaking of time-reversal symmetry is
intermediate. In particular, for small $\alpha$ one expects that
superconductivity should survive albeit with a reduced transition
temperature. Thus, $\alpha$ plays the role of a dimensionless pair-breaking
parameter that characterizes the relative importance of time-reversal
symmetry breaking disorder.

To proceed, we use the replica trick\cite{Edwards1975} to perform
the disorder average
\begin{equation}
\overline{e^{-S_{{\rm rdm}}}}=\overline{e^{-\sum_{ijk}g_{ijk}O_{ijk}}},
\end{equation}
where $O_{ijk}=\frac{1}{N}\sum_{\sigma a}\int_{0}^{\beta}d\tau c_{i\sigma a}^{\dagger}\left(\tau\right)c_{j\sigma a}\left(\tau\right)\phi_{ka}\left(\tau\right).$
Here, $a=1,\cdots,n$ is the replica index while $\tau$ stands for
the imaginary time in the Matsubara formalism with $\beta=\left(k_{B}T\right)^{-1}$
the inverse temperature. Using the Gaussian disorder distribution,
characterized by Eq.\eqref{eq:distribution}, we obtain for the disorder
average

\begin{equation}
\overline{e^{-\sum_{ijk}g_{ijk}O_{ijk}}}=e^{\frac{1}{4}\left(1-\frac{\alpha}{2}\right)\bar{g}^{2}\sum_{ijk}\left(O_{ijk}^{\dagger}+O_{ijk}\right)^{2}-\frac{\alpha}{8}\bar{g}^{2}\sum_{ijk}\left(O_{ijk}^{\dagger}-O_{ijk}\right)^{2}}.\label{eq:GOE}
\end{equation}

For the Gaussian unitary ensemble with $\alpha=1$ only terms like
$O_{ijk}^{\dagger}O_{ijk}$, occur. On the other hand, for $\alpha<1$ we find in
addition anomalous terms $O_{ijk}^{\dagger}O_{ijk}^{\dagger}$ and
$O_{ijk}O_{ijk}$. These terms give rise to anomalous self energies and
propagators and, if the self energies aquire a finite mean value, to
superconductivity.

To proceed, we introduce collective variables through the identities
\begin{eqnarray*}
1 & = & \int{\cal D}G{\cal D}\Sigma e^{\sum_{ab,\sigma\sigma'}\int d\tau d\tau'\left(NG_{ba,\sigma'\sigma}\left(\tau',\tau\right)-\sum_{i}c_{i\sigma a}^{\dagger}\left(\tau\right)c_{i\sigma'b}\left(\tau'\right)\right)\Sigma_{ab,\sigma\sigma'}\left(\tau,\tau'\right)},\\
1 & = & \int{\cal D}F{\cal D}\Phi e^{\sum_{ab,\sigma\sigma'}\int d\tau d\tau'\left(NF_{ba,\sigma'\sigma}\left(\tau',\tau\right)-\sum_{i}c_{i\sigma a}\left(\tau\right)c_{i\sigma'b}\left(\tau'\right)\right)\Phi_{ab,\sigma\sigma'}^{+}\left(\tau,\tau'\right)},\\
1 & = & \int{\cal D}F^{*}{\cal D}\Phi^{^{*}}e^{\sum_{ab,\sigma\sigma'}\int d\tau d\tau'\left(NF_{ba,\sigma'\sigma}\left(\tau',\tau\right)-\sum_{i}c_{i\sigma a}^{\dagger}\left(\tau\right)c_{i\sigma'b}^{\dagger}\left(\tau'\right)\right)\Phi_{ab,\sigma\sigma'}^{+}\left(\tau,\tau'\right)},\\
1 & = & \int{\cal D}D{\cal D}\Pi e^{\frac{1}{2}\sum_{ab}\int d\tau d\tau'\left(ND_{ba}\left(\tau',\tau\right)-\sum_{i}\phi_{ia}\left(\tau\right)\phi_{ib}\left(\tau'\right)\right)\Pi_{ab}\left(\tau,\tau'\right)}.
\end{eqnarray*}
This allows one to integrate out the electron and phonon degrees of
freedom, which yields for replica symmetric solutions and singlet
pairing the effective action

\begin{eqnarray}
S & = & -N{\rm tr}\log\left(\hat{G}_{0}^{-1}-\hat{\Sigma}\right)+\frac{N}{2}{\rm tr}\log\left(D_{0}^{-1}-\Pi\right)\nonumber \\
 & - & 2N\int d\tau d\tau'G\left(\tau',\tau\right)\Sigma\left(\tau,\tau'\right)+\frac{N}{2}\int d\tau d\tau'D\left(\tau',\tau\right)\Pi\left(\tau,\tau'\right)\nonumber \\
 & - & N\int d\tau d\tau'\left(F\left(\tau',\tau\right)\Phi\left(\tau,\tau'\right)+F^{*}\left(\tau',\tau\right)\Phi^{*}\left(\tau,\tau'\right)\right)\nonumber \\
 & + & N\overline{g}^{2}\int d\tau d\tau'\left(\left|G\left(\tau,\tau'\right)\right|^{2}-\left(1-\alpha\right)F^{*}\left(\tau,\tau'\right)F\left(\tau',\tau\right)\right)D\left(\tau,\tau'\right).
\end{eqnarray}
Here,
\begin{equation}
\hat{\Sigma}\left(\tau,\tau'\right)=\left(\begin{array}{cc}
\Sigma\left(\tau,\tau'\right) & \Phi\left(\tau,\tau'\right)\\
\Phi^{*}\left(\tau,\tau'\right) & -\Sigma\left(\tau',\tau\right)
\end{array}\right)
\end{equation}
is a matrix in Nambu space. In the limit $N\rightarrow\infty$ the
integration over the collective variables can be performed through
the saddle-point method and we obtain time-translation invariant solutions.
After Fourier transformation to Matsubara frequencies the saddle-point
equations take the form of the Eliashberg equations:
\begin{eqnarray}
i\epsilon_{n}\left(1-Z\left(\epsilon_{n}\right)\right) & = & -\bar{g}^{2}T\sum_{n^{\prime}}\frac{D\left(\epsilon_{n}-\epsilon_{n^{\prime}}\right)i\epsilon_{n'}Z\left(\epsilon_{n^{\prime}}\right)}{\left(\epsilon_{n'}Z\left(\epsilon_{n'}\right)\right)^{2}+\Phi\left(\epsilon_{n'}\right)^{2}},\nonumber \\
\Phi\left(\epsilon_{n}\right) & = & \left(1-\alpha\right)\bar{g}^{2}T\sum_{n'}\frac{D\left(\epsilon_{n}-\epsilon_{n^{\prime}}\right)\Phi\left(\epsilon_{n^{\prime}}\right)}{\left(\epsilon_{n'}Z\left(\epsilon_{n'}\right)\right)^{2}+\Phi\left(\epsilon_{n'}\right)^{2}},\nonumber \\
\Pi\left(\nu_{n}\right) & = & -2\bar{g}^{2}T\sum_{n'}\left(G\left(\epsilon_{n'}+\nu_{n}\right)G\left(\epsilon_{n}\right)-\left(1-\alpha\right)F\left(\epsilon_{n'}+\nu_{n}\right)F\left(\epsilon_{n'}\right)\right),
\label{eq:Eliash_eqs}
\end{eqnarray}
where $\Sigma\left(\epsilon_{n}\right)=i\epsilon_{n}\left(1-Z\left(\epsilon_{n}\right)\right)$.
\begin{equation}
D\left(\nu_{n}\right)=\frac{1}{\nu_{n}^{2}+\omega_{0}^{2}-\Pi\left(\nu_{n}\right)}
\end{equation}
is the phonon propagator, while the electron propagator in Nambu space
is
$\hat{G}^{-1}\left(\epsilon_{n}\right)=\hat{G}_0^{-1}-\hat{\Sigma}\left(\epsilon_{n}\right)$
with $\hat{G}_0^{-1}(\epsilon_n)=i\epsilon_n\hat{1}$.
Here $\epsilon_{n}$=$\left(2n+1\right)\pi T$ and $\nu_{n}=2n\pi T$ are
fermionic and bosonic Matsubara frequencies, respectively.

One immediately observes that in the normal state, with $\Phi=F=0$,
the solution of this set of equations is independent of the pair-breaking
parameter $\alpha$. Before we analyze the dependence of superconductivity
on $\alpha$ we briefly summarize the regimes $\alpha=1$ and
$\alpha=0$ that were discussed already in Ref.\cite{Esterlis2019}.

For $\alpha=1$, time reversal symmetry is strongly broken and no
superconducting solution exists. In the phase diagram as a function
of temperature and dimensionless coupling constant $g=\bar{g}/\omega_{0}^{3/2}$, shown in Fig.\ref{fig:phase_diag}a,
we find three distinct regimes: i) for $T\gg T_f\sim g^{2}\omega_{0}$ interactions
are irrelevant and the systems behaves as free fermions and bosons. ii) for $T< T^*$ with $T^*\sim \omega_{0}\min\left(g^{2},g^{-2}\right)$
we find a fermionic self energy $\Sigma\left(\epsilon_{n}\right)=-c_{1}i\epsilon_{n}\left|g^{2}\omega_{0}/\epsilon_{n}\right|^{2\Delta}$.
For the dynamic part of the bosonic self energy, $\delta\Pi\left(\nu_{n}\right)\equiv\Pi\left(\nu_{n}\right)-\Pi\left(0\right)$,
we obtain $\delta\Pi\left(\nu_{n}\right)=c_{3}\left|\nu_{n}/\left(g^{2}\omega_{0}\right)\right|^{4\Delta-1}\omega_0^2$.
The renormalized phonon frequency $\omega_{r}^{2}=\omega_{0}^{2}-\Pi\left(0\right)$
behaves as $\omega_{r}^{2}=c_{2}\left(T/\left(g^{2}\omega_{0}\right)\right)^{4\Delta-1}\omega_0^2$. The numerical coefficients $c_1 \approx 1.155$,
$c_2 \approx 0.561$, and $c_3 \approx 0.710$ were determined in Ref.\cite{Esterlis2019}. The scaling dimension of the fermions
was found to be $\Delta\approx 0.4203$. This exponent  determines the
phonon dynamics via the anomalous Landau damping term
$\delta\Pi\left(\nu_{n}\right)$. Notice, for $g<1$ it holds that $T_f\sim T^*$.
In this regime an emergent conformal symmetry characterizes the low energy and
low temperature behavior.
iii) For $g>1$ there exists an intermediate temperature regime $g^{-2}<T/\omega_{0}<g^{2}$, i.e. $T^* < T <T_f$,
where the phonons are long-lived and have a renormalized frequency $\omega_{r}^{2}=\left(\frac{3\pi}{8}\right)^{2}T\omega_0/g^{2}$.
At the same time the fermionic self energy is impurity-like, $\Sigma\left(\epsilon_{n}\right)=-i{\rm sign}\left(\epsilon_{n}\right)\frac{8}{3\pi}g^{2}\omega_0$,
a behavior caused by the coupling of electrons to very soft, i.e.
almost classical, phonons.

For $\alpha=0$, superconductivity emerges at a critical temperature
$T_{c}^{0}\left(g\ll1\right)\approx0.16g^{2}\omega_{0}$, while for large $g$
the transition temperature saturates,
$T_{c}^{0}\left(g\gg1\right)\approx0.112\omega_{0}$.  In the strong coupling
regime, the weight of the Bogoliubov quasiparticles behaves as $Z_{B} \sim
g^{-2}$ while the ratio $\frac{2\Delta_0}{T_c}\approx11.46$ with
superconducting gap $\Delta_0$ and the transition temperature $T_c$ is significantly
larger then the BCS value. We summarize the key regimes of the normal state and
the transition temperature in Fig.\ref{fig:phase_diag}a. The fact that
$T_c(g\ll1,\alpha=0)$ saturates is a consequence of the Anderson theorem where
impurities and soft bosons that do not break time reversal symmetry will not
cause a suppression of the transition temperature.

\section{Pair Breaking}

\subsection{Numerical analysis and gap equations in the scaling limit}

\begin{figure}
    \centering
\includegraphics{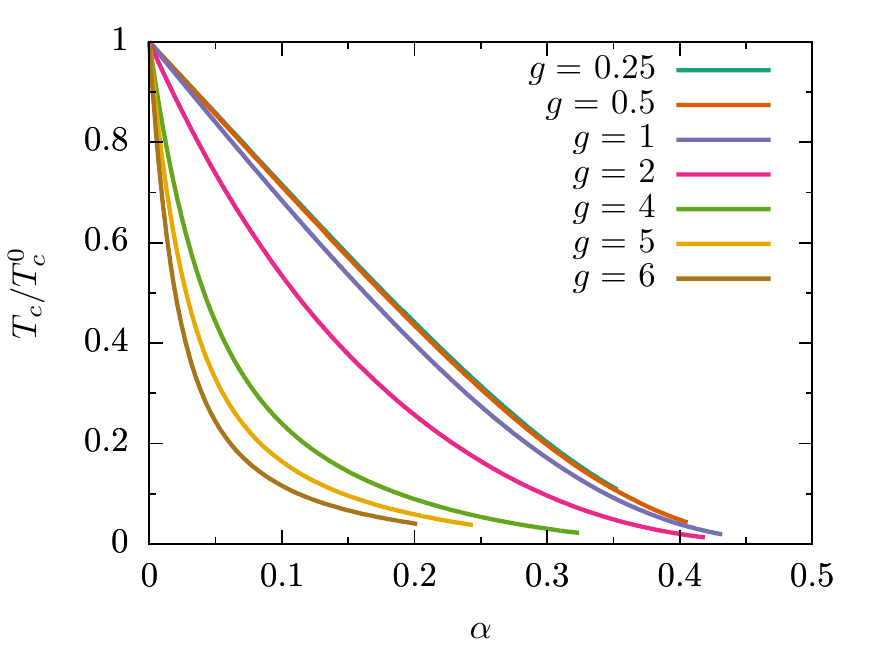}

\caption{Superconducting transition temperature $T_{c}\left(\alpha\right)$
as function of the dimensionless pair breaking parameter for different
dimensionless coupling strength $g$. For small $g$, $T_{c}\left(\alpha\right)$
follows a weakly $g$-dependent universal curve, while the initial
drop with the pair breaking strength depends sensitively on the coupling
constant for larger $g$. All curves seem to extend to a $g$-independent
critical value $\alpha_{c}$. The full numerical solution converges
poorly at low $T$ and large $g$, a regime that will be analyzed
by solving the linearized gap equation using analytically determined
scaling solutions for the normal-state electron and phonon propagators.}

\label{fig:full_numerics}
\end{figure}

\begin{figure}
    \centering
\includegraphics[]{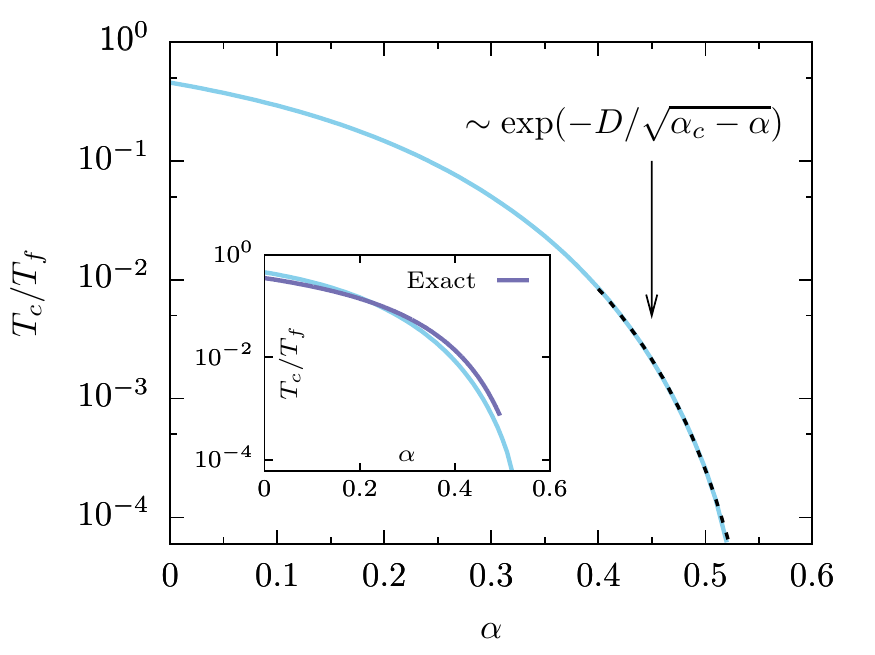}

\caption{Superconducting transition temperature normalized by $T_f$ as function of the pair-breaking strength determined from the solution of the gap equation Eq.\eqref{eq: lingap syk},
in comparison BKT-scaling behavior of Eq.\eqref{eq:BKTscaling} (dashed curve) with $\alpha_c \approx 0.62 $ and $D \approx 4.8$. Inset shows comparison of $T_c$ determined by solving Eq.\eqref{eq: lingap syk} with the exact solution for $g=1$, obtained by solving the full Eliashberg equations.}

\label{fig:bkt_scaling}
\end{figure}

We first present the results obtained from a complete numerical solution of the
coupled Eliashberg equations. In Fig.\ref{fig:full_numerics} we show the
superconducting transition temperature as a function of the pair-breaking
parameter $\alpha$ for varying dimensionless coupling strength $g$. While
$T_{c}\left(\alpha\right)$ is weakly $g$-dependent for small $g$, we find a
strong variation of the initial suppression of the transition temperature in
the strong coupling limit $g > 1$.  $T_{c}\left(\alpha\right)$ seems to vanish
at a $g$-independent critical value $\alpha_{c}$. While the numerical
convergence at large $g$ is poor at lowest temperatures, a behavior consistent
with Eq.\eqref{eq:BKTscaling} can clearly be seen for smaller $g$; see in
particular the inset of Fig.\ref{fig:bkt_scaling}.

In order to obtain a more detailed understanding of these results
we analyze the linearized gap equations in the scaling limit. If we
linearize with respect to the anomalous self energy and combine the
Eliashberg equations for $Z\left(\epsilon_{n}\right)$ and $\Phi\left(\epsilon_{n}\right)$
to determine the gap function $\Delta\left(\epsilon_{n}\right)=\Phi\left(\epsilon_{n}\right)/Z\left(\epsilon_{n}\right)$,
we obtain
\begin{equation}
\Delta\left(\epsilon_{n}\right)=\bar{g}^{2}T_{c}\sum_{n'}\frac{D\left(\epsilon_{n}-\epsilon_{n^{\prime}}\right)}{Z\left(\epsilon_{n'}\right)\epsilon_{n'}}\left(\left(1-\alpha\right)\frac{\Delta\left(\epsilon_{n'}\right)}{\epsilon_{n'}}-\frac{\Delta\left(\epsilon_{n}\right)}{\epsilon_{n}}\right).
\end{equation}
Here $D\left(\nu_{n}\right)=\left(\nu_{n}^{2}+\omega_{r}^{2}-\delta\Pi\left(\nu_{n}\right)\right)$
and $Z\left(\epsilon_{n}\right)=1-i\Sigma\left(\epsilon_{n}\right)/\epsilon_{n}$
are the self-consistently determined normal state boson propagator and fermion self energy. For
$\alpha=0$ the zeroth bosonic Matsubara frequency, i.e. $n=n'$,
does not contribute to the gap equation. This ensures that only quantum
fluctuations of phonons influence the value of $T_{c}$. Even very
soft bosons do not act as pair-breakers, in accordance with Anderson's
theorem \cite{Anderson1959,Abrikosov1958,Abrikosov1959}. The situation changes for finite $\alpha$. Now the zeroth
Matsubara frequency contributes to the gap equation and soft bosons
become pair breaking.

For $T<T^*$, the normal state boson propagator and fermion self energy
are governed by the SYK-like scaling regime. Inserting these results,
we obtain the following expression for the linearized gap equation:
\begin{equation}
\Delta\left(\epsilon_{n}\right)=\frac{1}{2\pi C_{\Delta}}\sum_{n'}\frac{\left(\frac{T_{f}}{T_{c}}\right)^{2\Delta}}{m_{0}+\left|n-n'\right|^{4\Delta-1}}\frac{\left(1-\alpha\right)\frac{\Delta\left(\epsilon_{n'}\right)}{n'+\frac{1}{2}}-\frac{\Delta\left(\epsilon_{n}\right)}{n+\frac{1}{2}}}{\left(n'+\frac{1}{2}\right)\left(\left(\frac{T_{c}}{T_{f}}\right)^{2\Delta}+\left|n'+\frac{1}{2}\right|^{-2\Delta}\right)}.\label{eq: lingap syk}
\end{equation}
where $m_{0}=\frac{c_{2}}{c_{3}\left(2\pi\right)^{4\Delta-1}}\approx0.157$
and $C_{\Delta}=-8\cos\left(\pi\Delta\right)\sin^{3}\left(\pi\Delta\right)\Gamma\left(2\Delta\right)^{2}\Gamma\left(1-4\Delta\right)/\pi^{2}$
such that $\frac{1}{2\pi C_{\Delta}}\approx0.168212$. Here
$T_{f}=\frac{1}{2\pi}c_{1}^{\frac{1}{2\Delta}}g^{2}\omega_{0}\approx0.189g^{2}\omega_{0}$.
Because the temperature only appears in the combination $T/T_f$, in this regime
we must have $T_c = T_f F(\alpha)$ for some function $F$. Furthermore, in the
weak-coupling regime $g \ll 1$ we have $T_f \sim T_c^0 \sim g^2$, so that
$T_c/T_c^0$ is independent of $g$. This explains the $g$-independence of
$T_c(\alpha)$ in the weak-coupling regime, observed in
Fig.\ref{fig:full_numerics}.  Gap equations similar to the one of Eq.\eqref{eq: lingap syk} were recently discussed in Ref.\cite{Wang2016,Abanov2019,Wu2019}, pointing out that superconductivity remains robust despite the incoherent nature of the normal state because the self-energy from dynamic critical fluctuations vanishes for the two lowest fermionic Matsubara frequencies.

 In Fig.\ref{fig:bkt_scaling} we show the dependence of the
transition temperature on $\alpha$, obtained from an analysis
of Eq.\eqref{eq: lingap syk}.  We also show a fit of $T_{c}\left(\alpha\right)$
to the BKT-scaling behavior of Eq.\eqref{eq:BKTscaling} that yields
a critical value $\alpha_{c}\approx0.623 \pm 0.003$, the coefficient $D\approx4.8\pm  0.1$, and $\log\left(T^*/T_f\right)=5.3\pm0.2$. In the next section we present an analytic analysis for $\alpha_c$ and $D$ that agrees well with these results.

%\begin{figure}
%\includegraphics[scale=0.2]{tc_loglog_fit}
%
%\caption{Superconducting transition temperature as function of the pair-breaking
%strength determined from the solution of the gap equation Eq.\eqref{eq: lingap syk}
%in comparison with the BKT-scaling behavior of Eq.\eqref{eq:BKTscaling}. }
%
%\end{figure}

\begin{figure}
    \centering
    \includegraphics[width=\linewidth]{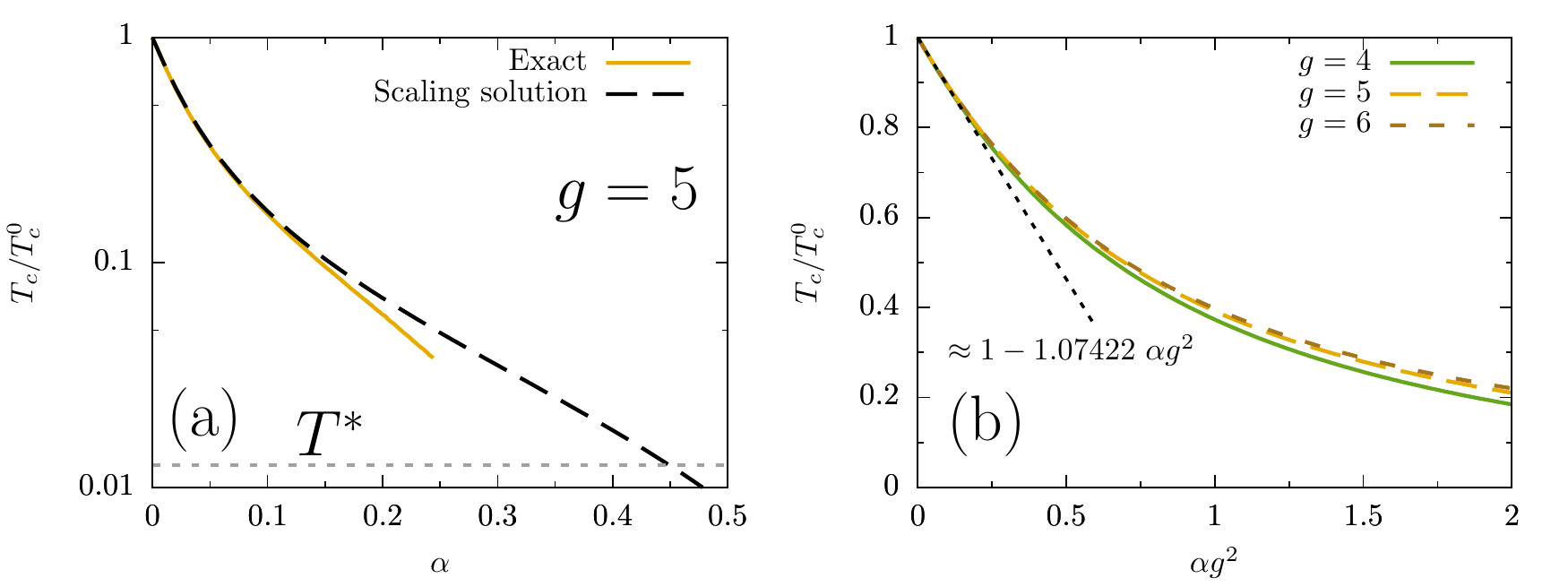}

\caption{(a) Comparison of the transition temperature obtained using the scaling solution in the strong coupling regime, Eq.\eqref{eq:lgap strong c}, with the full numerical solution for $g=5$. Dashed horizontal line is the crossover temperature $T^*$ below which the system enters the SYK-like regime. (b) The collapse of the transition temperature for large coupling constant $g$ is a function of $\alpha g^2$, with a linear slope of order unity. The data shown were obtained using the full numerical solution of the Eliashberg equation.}

\label{fig:tc_strong_coupling}
\end{figure}

In the impurity-like regime $T^*<T<T_f$ for  strong coupling $g>1$, the $\alpha=0$  transition temperature
approaches a constant value $T_{c}^{0}\approx0.112\omega_{0}$. For the crossover temperature $T^*$ we find $T^{*}=9\omega_0g^{-2}/256$.
As long as $T_{c}\left(\alpha\right)>T^{*}$
we can analyze the effects of pair breaking by solving the linearized
gap equation and using the normal state results from the impurity-like regime:
\begin{equation}
\Delta\left(\epsilon_{n}\right)=\frac{3\omega_{0}^{2}}{\left(8\pi T_{c}\right)^{2}}\sum_{n'}\frac{\left(1-\alpha\right)\frac{\Delta_{n'}}{n'+\frac{1}{2}}-\frac{\Delta_{n}}{n+\frac{1}{2}}}{\frac{T^{*}}{T_{c}}+\left(n-n'\right)^{2}}{\rm sign}\left(n'+\frac{1}{2}\right).\label{eq:lgap strong c}
\end{equation}
The solution of this equation is shown in Fig.\ref{fig:tc_strong_coupling}, in comparison with
the full numerical solution of Eq.\eqref{eq:Eliash_eqs}. The rapid drop of $T_c$ with $\alpha$ is well captured by
Eq.\eqref{eq:lgap strong c}. Of course, once $T<T^{*}$ Eq.\eqref{eq:lgap strong c} is no longer reliable, which we indicate by the horizontal dashed
line in Fig.\ref{fig:tc_strong_coupling}. The initial drop can be understood if one performs a perturbation
theory of the above matrix equation. Treating the gap equation as
an eigenvalue equation and performing first-order perturbation theory
in $\alpha$, ones finds that the perturbation is essentially diagonal,
being dominated by the zero Matsubara frequency transfer term in $D\left(\epsilon_{n}-\epsilon_{n'}\right)$,
and hence goes as $\omega_{0}/T^{*}\sim g^{2}$. Denoting the (normalized) $\alpha=0$ solution of the gap equation by $\Delta^{0}\left(\epsilon_{n}\right)$, to first order we have
\begin{equation}
\frac{T_{c}\left(\alpha\right)}{T_{c}^{0}}\approx1-\frac{4}{3\pi^{2}}\frac{\alpha g^{2}\omega_{0}}{T_{c}^{0}}\frac{\sum_{n}\frac{|\Delta^{0}_{n}|^{2}}{|2n+1|^{2}}}{\sum_{n}\frac{|\Delta^{0}_{n}|^{2}}{|2n+1|}}.
\end{equation}
The $\alpha=0$ solution can easily be determined numerically, from which we obtain
\begin{equation}
T_{c}\left(\alpha\right) \approx T_{c}^{0}\left(1-A ~ \alpha g^{2}\right)+{\cal O}\left(\alpha^{2}\right),
\end{equation}
with numerical coefficient $A\approx 1.07422$ close to unity. Thus, the transition temperature
is significantly reduced already for $\alpha\approx\alpha^{*}$ with
$\alpha^{*}\sim g^{-2}$. The reason for this behavior is the fragility
of a superconducting state with reduced weight of the Bogoliubov quasiparticles,
$Z_{B}\sim\frac{\omega_{0}}{T^{*}}\sim g^{-2}$. Thus, pair-breaking
effects are strong when   $\alpha$ becomes
comparable to the weight of the Bogoliubov quasi-particles, confirming
Eq.\eqref{eq:strong coupling Tc'} given above.

In the limit of large $g$ one can even go beyond the leading contribution
for small $\alpha$. To this end we split the sum in Eq.\eqref{eq:lgap strong c}
into the terms with $n\neq n'$ and $n=n'$:
\begin{eqnarray}
\Delta\left(\epsilon_{n}\right) & = & \frac{3\omega_{0}^{2}}{\left(8\pi T_{c}\right)^{2}}\sum_{n'\neq n}\frac{\left(1-\alpha\right)\frac{\Delta\left(\epsilon_{n'}\right)}{n'+\frac{1}{2}}-\frac{\Delta\left(\epsilon_{n}\right)}{n+\frac{1}{2}}}{\frac{T^{*}}{T_{c}}+\left(n-n'\right)^{2}}{\rm sign}\left(n'+\frac{1}{2}\right).\nonumber \\
 & - & \alpha g^{2}\frac{2\omega_{0}}{\pi^{2}T_{c}}\frac{\Delta\left(\epsilon_{n}\right)}{\left|n+\frac{1}{2}\right|}
\end{eqnarray}
Both terms contain the pair-breaking strength $\alpha$. However, in the
diagonal term $\alpha$ always enters in the combination
$\alpha g^{2}$. As this term dominates for large $g$, one expects that
$T_{c}\left(\alpha,g\right)\approx T_{c}\left(\alpha g^{2}\right)$
even beyond the leading order term in $\alpha$. This behavior is
verified in Fig.\ref{fig:tc_strong_coupling}.

\subsection{Analysis near $T=0$}

Finally we give qualitative arguments for the behavior near the quantum
critical point, including the BKT-scaling of the transition temperature
near $\alpha_{c}$, Eq.\eqref{eq:BKTscaling}. For $T\rightarrow0$
the SYK power-law behavior for the bosonic and fermionic propagators
properly describes the low-frequency dynamics regardless of the value
of the coupling constant $g$. This allows us to use the following
linearized gap equation
\begin{equation}
\Phi\left(\omega\right) =  \frac{1-\alpha}{C_{\Delta}}\int\frac{d\omega'}{2\pi}\frac{\Phi\left(\omega'\right)}{\left|\omega-\omega'\right|^{4\Delta-1}\left|\omega'\right|^{2-4\Delta}}.
\label{eq:zeroTgap}
\end{equation}
 The constant $C_{\Delta}$ was given above.

On the one hand we need to keep in mind  that the power-law behavior of this gap equation is only valid below the upper cutoff
  $T^* \approx\omega_{0}{\rm min}\left(g^{2},g^{-2}\right)$. In addition, we assume that we can use this $T=0$ gap equation to determine the transition temperature at very low $T$ if we introduce $T_c$ as lower cutoff. This yields:
\begin{equation}
\Phi\left(\omega\right)=\frac{1-\alpha}{2\pi C_{\Delta}}\int_{T_{c}}^{T^*}d\omega'\left(\frac{1}{\left|\omega-\omega'\right|^{4\Delta-1}}+\frac{1}{\left|\omega+\omega'\right|^{4\Delta-1}}\right)\frac{\Phi\left(\omega'\right)}{\omega'^{2-4\Delta}}.
\label{eq:zeroTgap2}
\end{equation}
In our subsequent analysis of Eq.\eqref{eq:zeroTgap2} we follow Ref.\cite{Abanov2001,Chubukov2005,Wang2016}.
Without the upper and lower cutoffs the
scale invariance of the problem suggests to look
for power-law solutions $\Phi\left(\omega\right)\sim\left|\omega\right|^{-b}$. As shown in  Ref.\cite{Abanov2001}, the natural solutions are of the form $b=\frac{4 \Delta-1}{2}\pm i \beta$, i.e. with finite imaginary part of the exponent, which yields
\begin{equation}
\Phi\left(\omega\right)\sim\left|\omega\right|^{-\frac{4\Delta-1}{2}}\frac{\sin\left(\beta\log\frac{T^{*}}{\left|\omega\right|}+\varphi\right)}{\sin \varphi}.
\label{solutionzeroTgap}
\end{equation}
The phase $\varphi$ is so far arbitrary and the  $\sin \varphi$ in the denominator was included for convenience. The imaginary part  $\beta$ of the exponent is determined by the implicit equation $1=\frac{1-\alpha}{2 \pi C_\Delta}J_\Delta(\beta)$ with
\begin{equation}
J_{\Delta}\left(\beta\right)=\int_{-\infty}^{\infty}dx\frac{1}{\left|1-x\right|^{4\Delta-1}\left|x\right|^{\frac{3-4\Delta}{2}+i\beta}}.
\end{equation}
The integral can be evaluated explicitly, but leads to somewhat lengthy expressions.
For small $\beta$ the implicit equation simplifies to $\alpha\approx \alpha_c-h \beta^2$, where $\alpha_c$ and $h$ depend on the exponent $\Delta$.
Inserting the value $\Delta \approx 0.4203$ we find $\alpha_{c} \approx 0.626531$ and $h\approx 2.37709$. Thus, solutions with real $\beta$ exists for $\alpha<\alpha_c$. We will now see that this is the critical pair-breaking strength where superconductivity disappears.

The effects of the infrared and ultraviolet cutoffs can  be incorporated via appropriate boundary conditions. Below $\omega =T_c$ the solution should be flat $\left.\frac{d\Phi\left(\omega\right)}{d\omega}\right|_{\omega=T_{c}}=0$. At the upper cut off, the natural condition for a power-law decay is that $ \left.\frac{d\Phi\left(\omega\right)}{d\omega}\right|_{\omega=T^{*}}=-(4 \Delta-1)\Phi\left(T^{*}\right)/T^{*}$. We will give a more detailed motivation of these conditions below. The ultraviolet boundary condition fixes the phase $\varphi$ via  $\cot \varphi=\frac{4 \Delta -1}{2 \beta}$. The infrared boundary condition then gives the condition for the transition temperature:
\begin{equation}
\beta \log \left(\frac{T^*}{T_c}\right)=\pi-2 \varphi.
\end{equation}
One sees that $T_c$ vanishes when $\beta$ vanishes. Using our above result near
the critical pair-breaking strength $\beta\approx
h^{-1/2}\sqrt{\alpha_{c}-\alpha}$, we obtain  the BKT-scaling behavior of
Eq.\eqref{eq:BKTscaling}
with $D= \pi \sqrt{h}\approx 4.843$. The analytic results for the critical pair-breaking strength $\alpha_c$ and for $D$ are in excellent agreement with our numerical results of Fig.\ref{fig:bkt_scaling}.
 The gap function  right at the critical point can be obtained by taking the $\beta \rightarrow 0$ limit:
\begin{equation}
\Phi\left(\omega\right)\sim\left|\omega\right|^{-\frac{4\Delta-1}{2}} \left(1+\frac{4 \Delta -1}{2} \log\frac{T^{*}}{\left|\omega\right|} \right).
\label{eq:powerlaw}
\end{equation}
The second, logarithmic term is a consequence of the boundary conditions at the upper cutoff. Right at the critical point, the overall amplitude of the gap is of course infinitesimal. The unusual frequency dependence can however be probed, at least in principle, if one measures the dynamic order-parameter susceptibility.

In order to get a better understanding of the above boundary conditions, we briefly summarize an alternative analysis that is controlled for
 $\Delta = 1/4+\epsilon$, with small $\epsilon$. This value for the exponent does indeed occur in
our model if we take the limit where the number of phonon modes $M$
is much larger than the number of fermion modes\cite{Esterlis2019},
in which case $\Delta=\frac{1}{4}+\sqrt{\frac{N}{8\pi M}}+\cdots$. Let us
consider without loss of generality $\omega>0$. We split the integration in
Eq.\eqref{eq:zeroTgap2} into regimes $\omega'<\omega$ and $\omega'>\omega$, which we approximate
as $\omega'\ll\omega$ and $\omega'\gg\omega$, respectively. Then it follows that
\begin{eqnarray}
\Phi\left(\omega\right) & \approx & \frac{1-\alpha}{C_{\Delta}\omega^{4 \Delta-1}}\int_{T_{c}}^{\omega}\frac{d\omega'}{\pi}\frac{\Phi\left(\omega'\right)}{\omega'^{2-4 \Delta}}+\frac{1-\alpha}{C_{\Delta}}\int_{\omega}^{\omega_{c}}\frac{d\omega'}{\pi}\frac{\Phi\left(\omega'\right)}{\omega'}.\label{eq:inteqapp}
\end{eqnarray}
For $\Delta \rightarrow \frac{1}{4}$ this approximation is exact, which makes the expansion controlled
for small $\epsilon$. This simplified version of the integral equation can be rewritten
as a differential equation. To this end we multiply Eq.\eqref{eq:inteqapp} with $\omega^{4\Delta-1}$, take the derivative with respect to $\omega$, multiply the result with $\omega^{2-4\Delta}$ and take once more a derivative. It follows:
\begin{equation}
\frac{d}{d\omega}\omega^{2-4\Delta}\frac{d}{d\omega}\omega^{4\Delta-1}\Phi\left(\omega\right)=-(1-\alpha)\frac{4\Delta-1}{\pi C_{\Delta}\omega}\Phi\left(\omega\right).
\label{diffeq}
\end{equation}
From the integral equation one furthermore obtains the  conditions $
T_{c}^{4\Delta-1}\Phi\left(T_{c}\right)=\frac{T_{c}}{4\Delta-1}\left.\frac{d\omega^{4\Delta-1}\Phi\left(\omega\right)}{d\omega}\right|_{\omega=T_{c}}$
    and
    $\left.\frac{d\omega^{\gamma}\Phi\left(\omega\right)}{d\omega}\right|_{\omega=\omega_{c}}
        =0$. These are identical to the boundary conditions imposed in our
        above analysis. In addition, the solution of Eq.\eqref{solutionzeroTgap}
        solves the differential equations Eq.\eqref{diffeq} with
        $\beta=\sqrt{\left(\frac{1-\alpha}{\pi
        C_{\Delta}}-\frac{1}{4}(4\Delta-1)\right)\left(4\Delta-1\right)}$. This
        yields for the critical pair-breaking strength
        $\alpha_{c}=\frac{1}{2}+\left(2\log4-\pi\right)\epsilon\cdots$.
We checked that this result also follows from the leading order expansion of the more general solution summarized above.

The  analysis of the preceding paragraphs clearly demonstrates that the
conformal symmetry of the critical normal state is lost with an energy
scale that behaves according to the BKT-scaling of Eq.\eqref{eq:BKTscaling},
in agreement with the general expectation of Ref.\cite{Kaplan2009}.

\section{Summary}

We introduced and solved a model of electrons that interact with
phonons via a random electron-phonon coupling. The
theory is a natural generalization of the Sachdev-Ye-Kitaev model
to the problem of interacting electrons and phonons and gives rise to a superconductivity. Typical for fully connected models, an exact solution becomes
possible in the limit $N\rightarrow\infty$. In our
case this exact solution corresponds to the coupled Eliashberg equations
of superconductivity. Since the normal state of the model is characterized
by two non-Fermi liquid fixed points, depending on the value of the
dimensionless coupling constant and temperature, the approach is an
ideal toy model to study superconductivity as it emerges from a quantum
critical non-Fermi liquid. In particular, a superconductor that results
from a strongly incoherent normal state is characterized by a reduced
weight $Z_{B}$ of the Bogoliubov quasi-particles.

The special focus of the present paper was the investigation of pair-breaking
phenomena. On the one hand, we found that superconductivity disappears
at a critical pair-breaking strength $\alpha_{c}$ according to BKT-scaling. This reflects the fact that the critical normal state possesses
at low temperatures an emergent conformal symmetry, which is broken
in the superconducting state. Thus, the way superconductivity is suppressed
by pair-breaking disorder may reveal important information about the
symmetry of the normal state. In addition we showed that for superconductors
with small weight $Z_{B}\ll1$ of the Bogoliubov quasiparticle, pair breaking
substantially suppresses $T_{c}$ already for
$\alpha\approx Z_{B}\ll\alpha_{c}$.

These results demonstrate that the formalism initially devised by Eliashberg
to treat dynamical pairing phenomena in Fermi liquids with intermediate
electron-phonon coupling is, in fact, general enough to address Cooper
pairing in a broader class of systems, such as strongly correlated,
quantum critical systems.

\section*{Acknowledgments}
We are grateful to Andrey V. Chubukov, Jonas Karcher, and Yoni Shattner for
helpful discussions. We are particularly thankful to Andrey V. Chubukov for
pointing out to us the importance of the logarithmic term in Eq.\eqref{eq:powerlaw}.


\begin{thebibliography}{10}
\bibitem{Eliashberg1960}G. M. Eliashberg, \emph{Interactions between
electrons and lattice vibrations in a superconductor}, Sov. Phys.
JETP \textbf{11}, 696 (1960).

\bibitem{Eliashberg1961}G. M. Eliashberg, \emph{Temperature Green's
functions for electrons in a superconductor}, Sov. Phys. JETP \textbf{12},
1000 (1961).

\bibitem{Migdal1958}A. B. Migdal, \emph{Interaction between electrons
and lattice vibrations in a normal metal}, Sov. Phys. JETP \textbf{7},
996 (1958).

\bibitem{Gorkov1958}L. P. Gorkov, \emph{On the energy spectrum of
superconductors}, Sov. Phys. JETP \textbf{34}, 505 (1958).

\bibitem{Nambu1960}Y. Nambu, \emph{Quasi-Particles and Gauge Invariance
in the Theory of Superconductivity}, Phys. Rev. \textbf{117}, 648
(1960).

\bibitem{Schrieffer1963}J. R. Schrieffer, D.J. Scalapino, and J.W.
Wilkins, E\emph{ffective tunneling density of states in superconductors},
Phys. Rev. Lett. \textbf{10}, 336 (1963)

\bibitem{Scalapino1966}D. J. Scalapino, J. R. Schrieffer, J. W. Wilkins,
\emph{Strong-coupling superconductivity}, Phys. Rev. \textbf{148},
263 (1966).

\bibitem{McMillan1968}W. L. McMillan, \emph{Transition Temperature
of Strong-Coupled Superconductors}, Phys. Rev. \textbf{167}, 331(1968).

\bibitem{Scalapino1969}D. J. Scalapino R.D. Parks (Ed.), \emph{Superconductivity,
The Electron\textendash Phonon Interaction and Strong Coupling Superconductors},
Vol. 1, Dekker Inc, New York (1969), p. 449

\bibitem{McMillan1969}W. L. McMillan, J. M. Rowell, M. Parks (Ed.),
\emph{Superconductivity, Tunneling and Strong Coupling Superconductivity},
Vol. 1, Dekker Inc, New York, p. 561 (1969).

\bibitem{Allen1975}P. B. Allen and R. C. Dynes, Transition temperature
of strong-coupled superconductors reanalyzed, Phys. Rev. B \textbf{12},
905 (1975).

\bibitem{Carbotte1990}J. P. Carbotte, P\emph{roperties of boson-exchange
superconductors}, Rev. Mod. Phys. \textbf{62}, 1027 (1990).

\bibitem{Bonesteel1996}N. E. Bonesteel, I. A. McDonald, and C. Nayak,
\emph{Gauge Fields and Pairing in Double-Layer Composite Fermion Metals},
Phys. Rev. Lett. \textbf{77}, 3009 (1996).

\bibitem{Son1999}D.T. Son, \emph{Superconductivity by long-range
color magnetic interaction in high-density quark matter}, Phys. Rev.
D \textbf{59}, 094019 (1999).

\bibitem{Abanov2001}Ar. Abanov, A. Chubukov, and A. Finkel\textquoteright stein,
\emph{Coherent vs. incoherent pairing in 2D systems near magnetic
instability}, Europhys. Lett. \textbf{54}, 488 (2001).

\bibitem{Abanov2001b}Ar. Abanov, A. V. Chubukov, and J. Schmalian,
\emph{Quantum-critical superconductivity in underdoped cuprates},
Europhys. Lett. \textbf{55}, 369 (2001).

\bibitem{Roussev2001}R. Roussev and A. J. Millis, \emph{Quantum critical
effects on transition temperature of magnetically mediated p-wave
superconductivity}, Phys. Rev. B \textbf{63}, 140504R (2001).

\bibitem{Chubukov2005}A. V. Chubukov and J. Schmalian, \emph{Superconductivity
due to massless boson exchange in the strong-coupling limit}, Phys.
Rev. B \textbf{72}, 174520 (2005).

\bibitem{Metlitski2015}M. A. Metlitski, D. F. Mross, S. Sachdev,
and T. Senthil, \emph{Cooper pairing in non-Fermi liquids}, Phys.
Rev. B \textbf{91}, 115111 (2015).

\bibitem{Raghu2015}S. Raghu, G. Torroba, and H. Wang, \emph{Metallic
quantum critical points with finite BCS couplings}, Phys. Rev. B \textbf{92},
205104 (2015).

\bibitem{Wang2016} Y. Wang, A. Abanov, B. L. Altshuler, E. A. Yuzbashyan, and A. V. Chubukov, \emph{Superconductivity near a Quantum-Critical Point: The Special Role of the First Matsubara Frequency},
Phys. Rev. Lett. \textbf{117}, 157001  (2016).

\bibitem{Abanov2019} A. Abanov, Y.-M. Wu, Y. Wang, and A. V. Chubukov, \emph{Superconductivity above a quantum critical point in a metal: Gap closing versus gap filling, Fermi arcs, and pseudogap behavior},
Phys. Rev. B \textbf{99}, 180506(R)  (2019).

\bibitem{Wu2019} Y.-M. Wu, A. Abanov, Y. Wang, and A. V. Chubukov, \emph{Special role of the first Matsubara frequency for superconductivity near a quantum critical point: Nonlinear gap equation below  $T_c$ and spectral properties in real frequencies},
Phys. Rev. B \textbf{99}, 144512 (2019).


\bibitem{Esterlis2019}I. Esterlis and J. Schmalian, \emph{Cooper
pairing of incoherent electrons: An electron-phonon version of the
Sachdev-Ye-Kitaev model}, Phys. Rev. B \textbf{100}, 115132 (2019).

\bibitem{Sachdev1993}S. Sachdev and J. Ye, \emph{Gapless spin liquid
ground state in a random, quantum Heisenberg magnet}, Phys. Rev. Lett.
\textbf{70}, 3339, (1993).

\bibitem{Georges2000}A. Georges, O. Parcollet, and S. Sachdev, \emph{Mean
Field Theory of a Quantum Heisenberg Spin Glass}, Phys. Rev. Lett.
\textbf{85}, 840 (2000).

\bibitem{Sachdev2010}S. Sachdev, \emph{Holographic Metals and the
Fractionalized Fermi Liquid}, Phys. Rev. Lett. \textbf{105}, 151602
(2010).

\bibitem{Kitaev2015}A. Kitaev, \emph{Hidden correlations in the Hawking
radiation and thermal noise}, Talk at KITP
\url{http://online.kitp.ucsb.edu/online/joint98/kitaev/},
February, 2015.

\bibitem{Kitaev2015b}A. Kitaev, \emph{A simple model of quantum holography}.
Talks at KITP http://online.kitp.ucsb.edu/online/entangled15/kitaev/
and http://online.kitp.ucsb.edu/online/entangled15/kitaev2/, April
and May, 2015.

\bibitem{Patel2018}A. A. Patel, M. J. Lawler, and E.-A. Kim, \emph{Coherent
Superconductivity with a Large Gap Ratio from Incoherent Metals},
Phys. Rev. Lett. \textbf{121}, 187001 (2018).

\bibitem{Wang2019}Y. Wang, A Solvable Random \emph{Model with Quantum-critical
Points for non-Fermi-liquid Pairing}, arXiv:1904.07240.

\bibitem{Chowdhury2019}D. Chowdhury and E. Berg, I\emph{ntrinsic
superconducting instabilities of a solvable model for an incoherent
metal}, arXiv:1908.02757

\bibitem{Anderson1959}P. W. Anderson, \emph{Theory of dirty superconductors},
J. Phys. Chem Solids \textbf{11}, 26 (1959).

\bibitem{Abrikosov1958}A. A. Abrikosov and L. P. Gor\textquoteright kov,
\emph{On the theory of superconducting alloys. 1. The electrodynamics
of alloys at absolute zero}, Zh. Eksp. Teor. Fiz. 35, 1558 (1958)
,{[}Sov. Phys. JETP 8, 1090 (1959){]}.

\bibitem{Abrikosov1959}A. A. Abrikosov and L. P. Gor\textquoteright kov,
\emph{Superconducting alloys at finite temperatures}, Zh. Eksp. Teor.
Fiz. 36, 319 (1959) ,{[}Sov. Phys. JETP 9, 220 (1959){]}.

\bibitem{Abrikosov1961}A. Abrikosov and L. P. Gor\textquoteright kov,
\emph{Contribution to the theory of superconducting alloys with paramagnetic
impurities,} Zh. Eksp. Teor. Fiz. 39, 1781 (1961) ,{[}Sov. Phys. JETP
12, 1243 (1961){]}.

\bibitem{ErwinMH1971}E. Müller-Hartmann and J. Zittartz, \emph{Kondo
effect in superconductors,} Phys. Rev. Lett. \textbf{26}, 428 (1971).

\bibitem{Yoksan1984}S. Yoksan and A. D. S. Nagi, \emph{Shiba-Rusinov
theory of magnetic impurities in anisotropic superconductors: Eliashberg
formalism,} Phys. Rev. B \textbf{30}, 2659 (1984).

\bibitem{Monthoux1994}P. Monthoux and D. Pines, \emph{Spin-fluctuation-induced
superconductivity and normal-state properties of YBa$_{2}$Cu$_{3}$O$_{7}$,}
Phys. Rev. B 49, 4261 (1994).

\bibitem{Preosti1996}G. Preosti and P. Muzikar, \emph{Superconducting
order parameters with sign changes: The density of states and impurity
scattering}, Phys. Rev. B 54, 3489 (1996).

\bibitem{Golubov1997}A. A. Golubov and I. I. Mazin, \emph{Effect
of magnetic and nonmagnetic impurities on highly anisotropic superconductivity,}
Phys. Rev. B \textbf{55}, 15146 (1997).

\bibitem{Franz1997}M. Franz, C. Kallin, A. J. Berlinsky, and M. I.
Salkola,\emph{ Critical temperature and superfluid density suppression
in disordered high- cuprate superconductors,} Phys. Rev. B \textbf{56},
7882 (1997).

\bibitem{Haran1998}G. Hara\'{n} and A. D. S. Nagi, \emph{Effect of
anisotropic impurity scattering in superconductors}, Phys. Rev. B
\textbf{58}, 12441 (1998).

\bibitem{Kulic1999}M. L. Kuli\'{c} and O. V. Dolgov, \emph{Anisotropic
impurities in anisotropic superconductors, }Phys. Rev. B \textbf{60},
13062 (1999).

\bibitem{Dzero2005}M. Dzero and J. Schmalian, \emph{Superconductivity
in Charge Kondo Systems,} Phys. Rev. Lett. \textbf{94}, 157003 (2005).

\bibitem{Balatsky2006}A. V. Balatsky, I. Vekhter, and J.-X. Zhu,
\emph{Impurity-induced states in conventional and unconventional }superconductors\emph{,
}Rev. Mod. Phys. \textbf{78}, 373 (2006).

\bibitem{Graser2007}S. Graser, P. J. Hirschfeld, L. Y. Zhu, and T.
Dahm, \emph{T$_{c}$ suppression and resistivity in cuprates with
out of plane defects,} Phys. Rev. B \textbf{76}, 054516 (2007).

\bibitem{Alloul2009}H. Alloul, J. Bobroff, M. Gabay, and P. J. Hirschfeld,
\emph{Defects in correlated metals and superconductors,} Rev. Mod.
Phys. \textbf{81}, 45 (2009).

\bibitem{Kemper2009}A. Kemper, D. G. S. P. Doluweera, T. A. Maier,
M. Jarrell, P. J. Hirschfeld, and H. P. Cheng, \emph{Insensitivity
of $d$-wave pairing to disorder in the high-temperature cuprate superconductors,}
Phys. Rev. B \textbf{79}, 104502 (2009).

\bibitem{Kogan2009}V. G. Kogan, \emph{Pair breaking in iron }pnictides\emph{,
}Phys. Rev. B \textbf{80}, 214532 (2009).

\bibitem{Vorontsov2009}A. B. Vorontsov, M. G. Vavilov, and A. V.
Chubukov, \emph{Superfluid density and penetration depth in the iron
pnictides}, Phys. Rev. B \textbf{79}, 140507R (2009).

\bibitem{Vorontsov2010}A. B. Vorontsov, Ar. Abanov, M. G. Vavilov,
and A. V. Chubukov, \emph{Reduced effect of impurities on the universal
pairing scale in the cuprates}, Phys. Rev B \textbf{81}, 012508 (2010).

\bibitem{Michaeli2012}K. Michaeli and L. Fu, \emph{Spin-Orbit Locking
as a Protection Mechanism of the Odd-Parity Superconducting State
against }Disorder\emph{, }Phys. Rev. Lett. \textbf{109}, 187003 (2012).

\bibitem{Hoyer2015}M. Hoyer, M. S. Scheurer, S. V. Syzranov, and
J. Schmalian, \emph{Pair breaking due to orbital magnetism in iron-based
superconductors,} Phys. Rev. B \textbf{91}, 054501 (2015).

\bibitem{Scheurer2015}M. S. Scheurer, M. Hoyer, and J. Schmalian,
\emph{Pair breaking in multiorbital superconductors: An application
to oxide interfaces,} Phys. Rev. B \textbf{92}, 014518, (2015).

\bibitem{Kang2016}J. Kang and R. M. Fernandes, \emph{Robustness of
quantum critical pairing against disorder}, Phys. Rev. \textbf{B 93,
224514 (2016).}

\bibitem{Trevisan2018}T. V. Trevisan, M. Schütt, and R. M. Fernandes,
\emph{Impact of disorder on the superconducting transition temperature
near a Lifshitz transition,} Phys. Rev. B \textbf{98}, 094514 (2018).

\bibitem{Berezinskii1972}V. L. Berezinskii, \emph{Destruction of
long range order in one-dimensional and two-dimensional systems having
a continuous symmetry group. I. Classical systems,} Sov. Phys. JETP,
32, 493 (1971); Sov. Phys. JETP, \textbf{34}, 610 (1972).

\bibitem{Kosterlitz1973}J. M. Kosterlitz and J. D. Thouless, \emph{Ordering,
metastability and phase transitions in two-dimensional }systems\emph{,
}Journal of Physics C: Solid State Physics, \textbf{6}, 1181 (1973).

\bibitem{Kaplan2009}D. B. Kaplan, J.-W. Lee, D. T. Son, and M. A.
Stephanov, \emph{Conformality lost}, Phys. Rev. D \textbf{80}, 125005 (2009).

\bibitem{Appelquist1988} T. Appelquist, D. Nash, and L. C. R. Wijewardhana, \emph{Critical Behavior in (2+1)-Dimensional QED}, Phys. Rev. Lett. \textbf{60}, 2575  (1988).

\bibitem{Mehta2004}M.L. Mehta, \emph{Random matrices}, 3rd edition,
Elsevier (2004).

\bibitem{Edwards1975}S. F. Edwards and P. W. Anderson, \emph{Theory
of spin glasses}, J. Phys. F \textbf{5}, 965 (1975).


\end{thebibliography}
\end{document}